\documentclass{PoS}

\newcommand{\Choose}[2]{{^{#1}C_{#2}}}
\def\abar{\overline{a}}
\usepackage{amsmath,amssymb}

\title{New developments in multi-meson systems}

\ShortTitle{New developments in multi-meson systems}

\author{\speaker{William Detmold}\\
       Thomas Jefferson National Accelerator Facility, 12000 Jefferson
       Ave, Newpot News, VA 23606, USA\\
        Department of Physics, College of William and Mary,
        Williamsburg, VA 23187, USA \\
       E-mail: \email{wdetmold@jlab.org}}

\author{Brian Smigielski\\
        Department of Physics, National Taiwan University, Taipei, Taiwan \\
        Department of Physics, College of William and Mary,
        Williamsburg, VA 23187, USA \\
        E-mail: \email{smigs@phys.ntu.edu.tw}}

      \abstract{New developments in the study of multi-meson systems
        are reviewed. We highlight a new recursive algorithm for
        generating the requisite contractions needed for studying
        complex systems of mesons involving large numbers of particles
        or multiple species of particles. First results on mixed
        species systems involving pions and kaons are also
        presented. }

\FullConference{The XXVIII International Symposium on Lattice Field Theory, Lattice2010\\
		June 14-19, 2010\\
		Villasimius, Italy}

\begin{document}

\section{Introduction}

Systems containing more than two mesons are of phenomenological
interest in a number of areas from heavy ion collisions at RHIC, to
the equation of state of neutron stars. In the last few years, there
has been a concerted effort by the NPLQCD collaboration to study such
systems from first principles using lattice QCD
\cite{Beane:2007es,Detmold:2008fn,Detmold:2008yn}.  Much progress has
been made and these studies enabled the first calculation of a three
hadron interaction, that between three like-charged pions.

On the theoretical side, a number of developments have also occurred
in the last few years. First, previous results for the volume
dependence of the ground state energies of $n$ boson systems
\cite{Beane:2007qr,Tan:2007bg,Detmold:2008gh} have been extended to
the case of mixed species systems involving pions and kaons
\cite{Smigielski:2008pa}. Secondly, a new algorithm has been presented
\cite{Detmold:2010au} that recursively generates the quark propagator
contractions for complicated systems from those of smaller systems. In
principle, this method allows the calculation of systems with very
large numbers of mesons (surmounting the limit of $n=12$ particles in
previous calculations) and of systems involving many different types
of bosons.
In this contribution, recent aspects of this progress are
highlighted.

\section{Multi-Meson Interactions}

\label{sec:npi}

It has long been known how to exploit the volume dependence of the
eigen-energies of two hadron systems to extract infinite volume
scattering phase shifts~\cite{Luscher:1986pf} provided that the
effective range of the interaction, $r$ is small compared to the
spatial extent of the lattice volume (since $r\sim m_\pi^{-1}$ for
most interactions, this constraint is equivalent to $m_\pi\ L
\gg1$). In recent works, this has been extended to systems involving
$n>2$ bosons~\cite{Beane:2007qr,Tan:2007bg,Detmold:2008gh} and $n=3$
fermions~\cite{Luu:2008fg} in the situation where the relevant
scattering length, $a$, is small compared to the spatial extent of the
lattice.  By performing a perturbative calculation involving two- and
three- body hadronic interactions in a finite volume, the ground state
energy of a system of $n$ bosons was computed in
Refs.~\cite{Beane:2007qr,Tan:2007bg,Detmold:2008gh}. The shift in
energy of $n$ bosons of mass $M$ from the non-interacting system is
\begin{eqnarray}
 \Delta E_n &=&
  \frac{4\pi\, \abar}{M\,L^3}\Choose{n}{2}\Bigg\{1
-\left(\frac{\abar}{\pi\,L}\right){\cal I}
+\left(\frac{\abar}{\pi\,L}\right)^2\left[{\cal I}^2+(2n-5){\cal J}\right]
\nonumber 
\\&&\hspace*{2cm}
-
\left(\frac{\abar}{\pi\,L}\right)^3\Big[{\cal I}^3 + (2 n-7)
  {\cal I}{\cal J} + \left(5 n^2-41 n+63\right){\cal K}\Big]
\nonumber
\\&&\hspace*{2cm}
+
\left(\frac{\abar}{\pi\,L}\right)^4\Big[
{\cal I}^4 - 6 {\cal I}^2 {\cal J} + (4 + n - n^2){\cal J}^2 
+ 4 (27-15 n + n^2) {\cal I} \ {\cal K}
\nonumber\\
&&\hspace*{4cm}
+(14 n^3-227 n^2+919 n-1043) {\cal L}\ 
\Big]
\Bigg\}
\nonumber\\
&&
+\ \Choose{n}{3}\left[\ 
{192 \ \abar^5\over M\pi^3 L^7} \left( {\cal T}_0\ +\ {\cal T}_1\ n \right)
\ +\ 
{6\pi \abar^3\over M^3 L^7}\ (n+3)\ {\cal I}\ 
\right]
\nonumber\\
&&
+\ \Choose{n}{3} \ {1\over L^6}\ \overline{\overline{\eta}}_3^L\ 
\ \ + \ {\cal O}\left(L^{-8}\right)
\ \ \ \ ,
\label{eq:energyshift}
\end{eqnarray}
where the parameter $\abar$ is related to the scattering length, $a$,
and the effective range, $r$, by
\begin{eqnarray}
a
& = & 
\overline{a}\ -\ {2\pi\over L^3} \overline{a}^3 r \left(\ 1 \ -\
  \left( {\overline{a}\over\pi L}\right)\ {\cal I} \right)\ \ .
\label{eq:aabar}
\end{eqnarray}
The geometric constants, ${\cal I},\ {\cal J},\ {\cal K},\ {\cal L},\,
{\cal T}_{0,1}$, that enter into Eq.~(\ref{eq:energyshift}) are
defined in Ref.~\cite{Detmold:2008gh} and $^nC_m$ are the binomial 
coefficients.  The
three-body contribution to the energy-shift given in
Eq.~(\ref{eq:energyshift}) is represented by the parameter
$\overline{\overline{\eta}}_3^L$ (see Ref.~\cite{Detmold:2008gh}).

Using lattice QCD calculations of these energy shifts, the parameters
$\abar$ and $\overline{\overline{\eta}}_3^L$ can be extracted (this
method was used in
Refs.~\cite{Beane:2007es,Detmold:2008fn,Detmold:2008yn}).  To
determine the energy shifts, the multi-meson correlation functions
(specifying to the multi-pion system)
\begin{eqnarray}
C_n(t) 
 & \propto & \left\langle 
\left(\sum_{\bf x} \pi^-({\bf x},t)
\right)^n
\left( 
\phantom{\sum_x\hskip -0.2in}
\pi^+({\bf 0},0)
\right)^n
\right\rangle\
 \ \ ,
\label{eq:Cnfun}
\end{eqnarray}
are calculated.
On a lattice of infinite temporal extent,\footnote{Effects of temporal
  (anti-)periodicity are discussed in Ref.~\cite{Detmold:2008yn}.} the
combination
\begin{eqnarray}
G_n(t) 
&  \equiv & { C_n(t) \over \left[\ C_1 (t)\ \right]^n }
\ \stackrel{t\to\infty}{\longrightarrow}\ {\cal B}_0^{(n)}\ e^{- \Delta E_n\ t}
\ \ \ ,
\label{eq:Gnlarget}
\end{eqnarray}
allows for an extraction of the ground-state energy shift, $\Delta E_n$, which can
then be used as input into Eq.~(\ref{eq:energyshift}) to extract the scattering
and interaction parameters.
To compute the $(n!)^2$ Wick contractions in Eq.~(\ref{eq:Cnfun}),
the correlation function can be written as
\begin{eqnarray}
C_n(t) 
 & \propto & 
\langle \ \left(\ \overline{\eta} \Pi \eta\ \right)^n \ \rangle
\ \propto \ 
\varepsilon^{\alpha_1\alpha_2..\alpha_n\xi_1..\xi_{12-n}}\ 
\varepsilon_{\beta_1\beta_2..\beta _n\xi_1..\xi_{12-n}}\ 
\left(\Pi\right)_{\alpha_1}^{\beta_1} \left(\Pi\right)_{\alpha_2}^{\beta_2} 
.. \left(\Pi\right)_{\alpha_n}^{\beta_n} 
\ \ \ ,
\nonumber\\
\Pi
& = &  \sum_{\bf x} \ S({\bf x},t;0,0) \   S^\dagger({\bf x},t;0,0)
 \ \ ,
\label{eq:Cnfungrassman}
\end{eqnarray}
where $S({\bf x},t;0,0)$ is a light-quark propagator.  
The object
(block) $\Pi$ is a $N\times N$  bosonic
time-dependent matrix where $N=12$ 
($N=N_S\times N_C$ with $N_S=4$-spin and $N_C=3$-color), 
and $\eta_\alpha$ is a twelve component
Grassmann variable. 
Further simplifications are possible resulting in the
correlation functions being written in terms of traces of powers of $\Pi$.
As an example, the
contractions for the $3$-$\pi^+$ system give
\begin{eqnarray}
C_3(t) & \propto & 
{\rm tr}\left[ \Pi \right]^3
\ -\  3\  {\rm tr}\left[ \Pi^2 \right] {\rm tr}\left[\Pi\right]
\ +\  2\  {\rm tr}\left[ \Pi^3 \right]
\ \ \ ,
\label{eq:threePiCorrelator}
\end{eqnarray}
where the traces are over color and spin indices.

\section{Contractions for Large ($N>12$) Systems of Mesons}
\label{sec:contr-large-n12}

To extend these types of calculations to systems of $n>12$ mesons, or
to study systems of many different types of mesons, different methods
of performing the contractions of quark fields are required. While
better than the na\"ive factorial construction (which scales as
$n!^2$), the construction used in Section~\ref{sec:npi} scales poorly
to large numbers of mesons, behaving at best as $n!^{1/2}$ (provided
the matrix $\Pi$ is generalized to give a nonzero result). In
Ref.~\cite{Detmold:2010au}, a recursive method for performing these
contractions was developed that allows the extension of the study of
meson systems to larger $n$ and also greatly simplifies the
contractions required for systems of many different species of
mesons. We will outline the construction by considering the recursive
approach to the contractions for a single species of meson before
reporting the general case.

By rescaling the correlation functions of the single-species, single-source
system considered previously as 
\begin{eqnarray}
  C_{n\pi^+}(t)   & = &   (-)^n\ n!\ \langle\ R_n\ \rangle
  \ \ \ \ ,
  \label{eq:npiDET}
\end{eqnarray}
(the angle brackets denote a trace over spin and color indices) it is
straightforward to show that the objects, $R_n$ that are defined
implicitly in the above equation satisfy an ascending
recursion 
\begin{eqnarray}
  R_{n+1} & = & \langle\ R_n\ \rangle\ \Pi\ -\ n\ R_n\ \Pi \, .
  \label{eq:npipRECURSION}
\end{eqnarray}
with the initial condition that $R_1=\Pi$ as defined in
Eq.~(\ref{eq:Cnfungrassman}).  To see how
this works we explicitly construct the first few terms:
\begin{eqnarray}
  R_2 & = & \langle\ R_1\ \rangle \Pi \ -\ R_1\ \Pi
  \ =\ \langle\ \Pi\ \rangle\ \Pi \ -\ \Pi^2
  \nonumber\\
  \langle \ R_2\ \rangle & = &
 \langle\ \Pi\ \rangle^2 \ -\ \langle\ \Pi^2\ \rangle
  \ \ ,
\nonumber\\
  R_3 & = & \langle\ R_2\ \rangle\  \Pi \ -\ 2\ R_2\ \Pi
  \ =\  
  \langle\ \Pi\ \rangle^2 \ \Pi
  \ -\ \langle\ \Pi^2\ \rangle\ \Pi
  \ -\ 2\ \langle\ \Pi\ \rangle\ \Pi^2 
  \ +\ 2\ \Pi^3\ 
  \nonumber\\
  \langle\ R_3 \ \rangle 
  & = & 
  \langle\ \Pi\ \rangle^3
  \ -\ 3\ \langle\ \Pi^2\ \rangle\ \  \langle\ \Pi\ \rangle\
  \ -\ 2\ \langle\ \Pi^3\ \rangle\ 
  \ \ ,
  \label{eq:pipR3}
\end{eqnarray}
in agreement with Eq.~(\ref{eq:threePiCorrelator}).
Descending recursions also exist.

A correlation function for a system composed of $n_{ij}$
mesons of the $i^{\rm th}$ species from the $j^{\rm th}$ source at
$({\bf y}_j,0)$, where $0\le i\le k$ and $0\le j\le m$, is of the form
\begin{eqnarray}
  && C_{\bf n}(t) \ = \  
  \Bigg\langle\ 
  \left(\ \sum_{\bf x}\ {\cal A}_1 ({\bf x},t)\ \right)^{{\cal N}_1} \ 
  ...
  \left(\ \sum_{\bf x}\ {\cal A}
    _k ({\bf x},t)\ \right)^{{\cal N}_k} \
  \nonumber\\
  &&
  \left( \phantom{\sum_{\bf x}}\hskip -0.22in
    {\cal A}_1^\dagger({\bf  y_1},0) \right)^{n_{11}} ...\  
  \left( \phantom{\sum_{\bf x}}\hskip -0.22in
    {\cal A}_1^\dagger({\bf  y_m},0) \right)^{n_{1m}} ...\  
  \left( \phantom{\sum_{\bf x}}\hskip -0.22in
    {\cal A}_k^\dagger({\bf  y_1},0) \right)^{n_{k1}} ...\  
  \left( \phantom{\sum_{\bf x}}\hskip -0.22in
    {\cal A}_k^\dagger({\bf  y_m},0) \right)^{n_{km}}  
  \Bigg\rangle
  \ ,
  \label{eq:mk}
\end{eqnarray}
where ${\cal N}_i = \sum_j\ n_{ij}$ is the total number of mesons of
species $i$, and the subscript in $C_{\bf n} (t)$ labels the number of
each species from each source,
\begin{eqnarray} {\bf n} & = &\left(
    \begin{array}{cccc}
      n_{11}&n_{12}&...&n_{1m}\\
      \vdots &\vdots &\vdots &\vdots \\
      n_{k1} & n_{k2}&...&n_{km}
    \end{array}
  \right)
  \ \ \ .
  \label{eq:nvecdef}
\end{eqnarray}
The ${\cal A}_i({\bf y},t)$ denotes a quark-level interpolating 
operator ${\cal A}_m
({\bf x},t) = \overline{q}_m({\bf x},t) \ \gamma_5\ u ({\bf x},t)$, and
it can be shown that~\cite{Detmold:2010au} 
\begin{eqnarray}
  && C_{\bf n}(t) \ = \  
  \left(\ \prod_i\ {\cal N}_i!\ \right)\ 
  \left\langle\ \prod_{i,j} 
  \left(\ \overline{\eta}\ P_{ij}\ \eta\ \right)^{n_{ij}}\ \right\rangle
\ =\ 
  (-)^{\overline{\cal N}}\ 
  {\left(\ \prod_i\ {\cal N}_i!\right)\  \ \left(\ \prod_{i,j}\ n_{ij}!
    \right)\ 
    \over \overline{\cal N}!}
  \ 
  \langle\ T_{  {\bf n } }\ \rangle
  \ \ \ ,
  \label{eq:mkGRASS}
\end{eqnarray}
where the $\eta$ are $m\times N$-component Grassmann variables, and
the $P_{ij}$ are $\overline{N}\times\overline{N}$ dimensional
matrices, where $\overline{N}=m\times N$, which are generalizations of
the $\Pi$ defined in Eq.~(\ref{eq:Cnfungrassman}) with an additional
species index, $i$. They are defined as
\begin{eqnarray}
  P_{ij} & = & 
  \left(
    \begin{array}{c|c|c|c}
      0&0&...&0 \\
      \hline
      \vdots & \vdots & ... & \vdots \\
      \hline
      \left(A_i\right)_{j1}(t)&\left(A_i\right)_{j2}(t)&\ \  ... \ \ & \left(A_i\right)_{jm}(t) \\
      \hline
      0&0& ... & 0\\
      \hline
      \vdots & \vdots & ... & \vdots \\
      \hline
      0&0& ... & 0
    \end{array}
  \right)
\ \ ,
\nonumber\\
  \left(\ A_i\ \right)_{ab} & = & 
  \sum_{\bf x}\ S({\bf x},t;{\bf y}_b,0)\ S_i^\dagger ({\bf x},t;{\bf
    y}_a,0)\ 
  \ \ \ ,
  \label{eq:Sijdef}
\end{eqnarray}
The $ \left(\ A_i\ \right)_{ab}$
are $N\times N$ dimensional matrices, one for each flavor, $i$, and
pair of source indices, $a$ and $b$.  
$\overline{\cal N} = \sum_i\ {\cal N}_i$ is the total number of
mesons in the system, with $\overline{\cal N} \le \overline{N}$.  The
$T_{{\bf n} }$ 
defined implicitly in Eq.~(\ref{eq:mkGRASS})
satisfy the recursion relation
\begin{eqnarray}
  T_{  {\bf n}+{\bf 1}_{rs} }
  & = & 
  \sum_{i=1}^k\ 
  \sum_{j=1}^m\ 
  \ 
  \langle\ T_{ {\bf n}  + {\bf 1}_{rs} - {\bf 1}_{ij}  }\ \rangle\ P_{ij}
  \ -\
  \overline{\cal N}\  T_{  {\bf n}  + {\bf 1}_{rs} - {\bf 1}_{ij} }\ \ P_{ij}
  \ \ \ \ ,
  \label{eq:genRECUR}
\end{eqnarray}
where 
\begin{eqnarray} 
{\bf 1}_{ij} \ & = &\left(
    \begin{array}{cccc}
      0 & 0& \cdots & 0\\
      \vdots & \vdots  & \hdots \  1 \ \hdots & \vdots \\
      0 & 0&\cdots&0
    \end{array}
  \right)
  \ \ \ ,
  \label{eq:nplusvecdef}
\end{eqnarray}
and where the sole non-zero value is in the $(i,j)^{\rm th}$ entry.
Defining ${\cal U}_j = \sum_i \ n_{ij}$ to be the number of mesons
from the $j^{\rm th}$ source, it is clear that the correlation
function vanishes when ${\cal U}_j>N$ for any source $j$.

These recursion relations (and those for the simpler systems also
studied in Ref.~\cite{Detmold:2010au}) allow for the calculation of
arbitrarily large systems of mesons.  As formulated above, the systems
are restricted to contain quarks of one flavor but anti-quarks of any
number of flavors~\footnote{This restriction is not necessary, but
  relaxing it results in much more complex sets of recursions.} or
vice versa.  Importantly the above algorithm scales linearly with the
total number of mesons in the system.

This recursive construction has been implemented and is currently
being used to study a variety of complex meson systems that were
previously not accessible. Generalisations to the case of baryons are
straightforward but cumbersome.

\section{Mixed Species Meson Systems}
\label{sec:mixed-species-meson}

In the case of a mixed system comprising $n$ mesons of one type and
$m$ mesons of a second type, the results summarized in
Section~\ref{sec:npi} have been generalized in
Ref.~\cite{Smigielski:2008pa}, working to ${\cal O}(1/L^6)$. The
energy shift of the interacting system from the free system depends on
three two-body interaction parameters and four three-body interaction
parameters. As might be expected, the full form of the energy shift is
lengthy and we refer the interested reader to the original paper.

These systems are currently under numerical study and preliminary
results using $L^3\times T=20^3\times128$ anisotropic clover lattices
at a pion mass of 390 MeV generated by the Hadron Spectrum
Collaboration are discussed herein.

In order to extract the energies of the multi meson systems, we  study
the behavior of two point correlation functions
 \begin{equation}
C_{N,M}(t)=\Big\langle \Big(\sum_{\bf{x}} \pi^{-}(\bf{x},t) \Big)^N
\Big( \sum_{\bf{x}} K^{-}(\bf{x},t) \Big)^M  \Big(\pi^{+}(\bf{0},t)
\Big)^N \Big(K^{+}(\bf{0},t) \Big)^M  \Big\rangle\,,
\label{eq:Cnm}
\end{equation}
where the pion and kaon interpolating operators are defined in terms
of quark fields as $\pi^+({\bf x},t)=\overline{u}({\bf
  x},t)\gamma_5d({\bf x},t)$ and $K^+({\bf x},t)=\overline{u}({\bf
  x},t)\gamma_5d({\bf x},t)$, respectively.

At large Euclidean times, $0\ll t\ll T$, these correlation functions
decay exponentially with the ground state energies, $E_{N,M}$, of the
system. These energies could then be used to constrain the two- and
three- body interactions using the results of
Ref.~\cite{Smigielski:2008pa}. However, satisfying the constraint
$0\ll t\ll T$, requires taking the limit of vanishing temperature and
is not typically practical in current lattice calculations. The
presence of thermal fluctuations that vanish exponentially as $T\to
\infty$, leads, at finite temporal extent, to significantly more
complex behaviour of the correlation functions in Eq.~(\ref{eq:Cnm}).
The general form for these correlation functions can be
straightforwardly deduced from Hamiltonian evolution at finite
temperature and is given by the following
\begin{eqnarray}
C_{N,M}(t)&=&\frac{1}{2}\sum_{m=0}^M \sum_{n=0}^N Z^{N-n,M-m}_{n,m}
e^{- \left(E_{N-n,M-m}+E_{n,m}\right)T/2} \cosh\left(
  \Big(E_{N-n,M-m}-E_{n,m}\Big) (t-T/2) \right) \nonumber \\  
&& +
\frac{1}{2}Z^{\frac{N}{2},\frac{M}{2}}_{\frac{N}{2},\frac{M}{2}} \
e^{\left(E_{N/2,M/2}\right)T/2} \ \delta_{N,2l} \delta_{M,2k}  +\ldots
\label{eq:Cnmhadronic}
\end{eqnarray}
where the last term is only present when $N$ and $M$ are both even
numbers. The ellipsis denotes excited state contributions. The
subleading terms shown in the summation correspond to the various ways
one can connect the $N$-pion, $M$-kaon source and sink, allowing for
propagation of subsets of hadrons around the temporal boundary
(various of the $Z_{n,m}^{N,M}$ are simply related). 

\begin{figure}[!t]
  \centering
  \includegraphics[width=0.7\columnwidth]{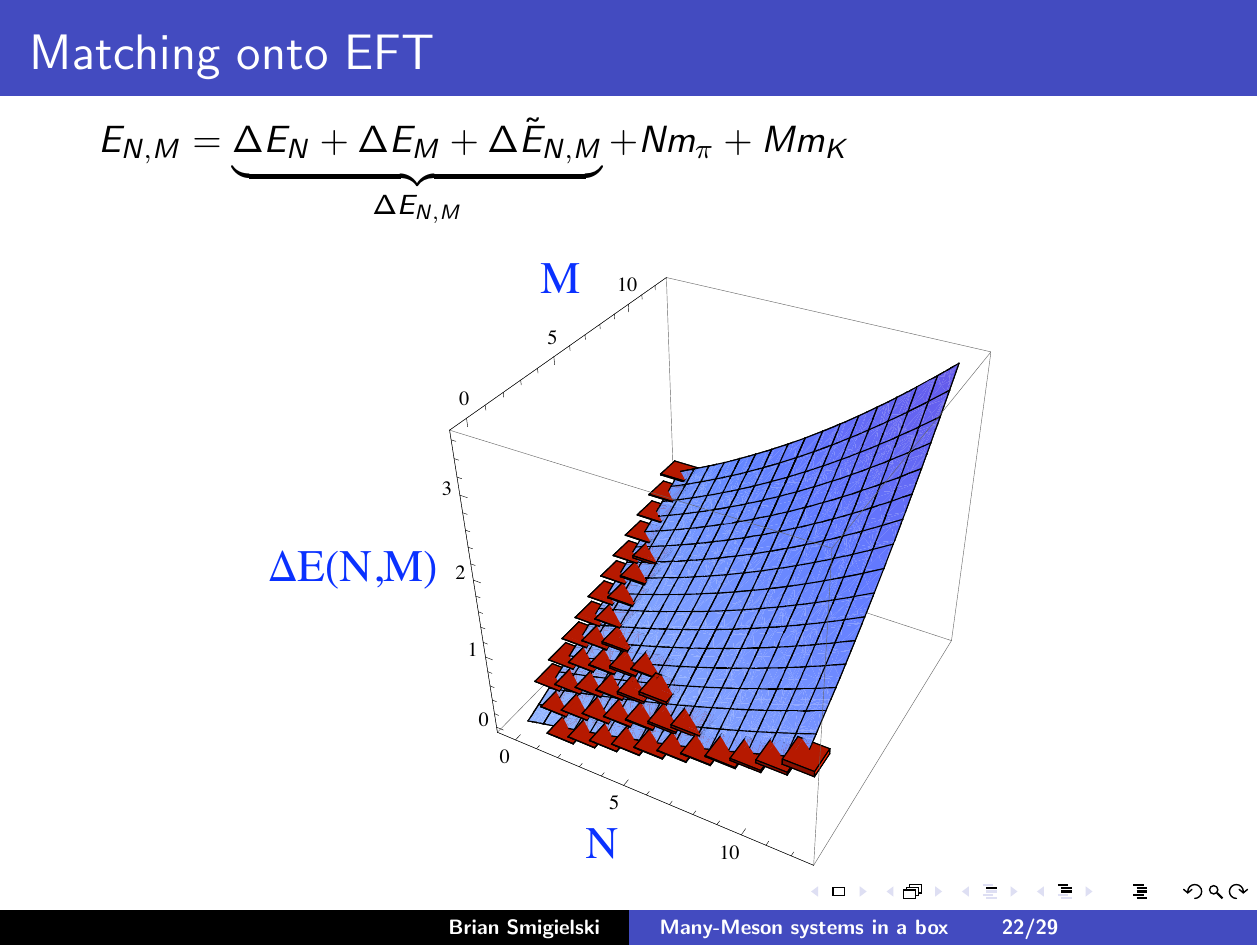}
  \caption{Energies of systems of $N$ pions and $M$ kaons extracted
    from our calculations. The vertical extent of the shaded
    boxes corresponds the uncertainty in the determination of the
    energies.}
  \label{fig:nm}
\end{figure}

With the large number of parameters that occur in
Eq.~(\ref{eq:Cnmhadronic}), fitting the correlation functions becomes
a complicated endeavour. Our current analysis strategy employs
variable projection (VarPro \cite{Kaufman:1975,Fleming:2004hs}) to
eliminate the linear parameters, the $Z^{N-n,M-m}_{n,m}$, from the
minimisation procedure. As $N$ and $M$ increase, we also employ the
method of Bayesian priors \cite{Lepage:2001ym} by providing suitable
physical starting points for the value of the energies in the
correlation function. A summary plot of the energy extractions is
shown in Fig.~\ref{fig:nm}.

Analysis of these correlation functions is ongoing and preliminary
results are very encouraging. The data provide statistically significant
extractions of most of the two- and three- body interactions that
enter the mixed-species systems. It additionally allows us to study
the relationships between the isospin and hypercharge densities and
the corresponding chemical potentials.

\section{Summary}
\label{sec:summary}

Systems of large numbers of mesons provide a novel form of QCD matter
that is amenable to study using lattice QCD. A rich phase structure is
expected and the tools and methodologies presented herein will allow a
close study of these systems. These techniques also represent an
important step towards the complexity frontier that must be confronted
if lattice QCD is to be applied to the physics of nuclei.



\begin{thebibliography}{99}

\bibitem{Beane:2007es}
  S.~R.~Beane, W.~Detmold, T.~C.~Luu, K.~Orginos, M.~J.~Savage and A.~Torok,
  Phys.\ Rev.\ Lett.\  {\bf 100}, 082004 (2008)
  [arXiv:0710.1827 [hep-lat]].

\bibitem{Detmold:2008fn}
  W.~Detmold, {\it et al.}, 
  Phys.\ Rev.\  D {\bf 78}, 014507 (2008).

\bibitem{Detmold:2008yn}
  W.~Detmold, K.~Orginos, M.~J.~Savage and A.~Walker-Loud,
  Phys.\ Rev.\  D {\bf 78}, 054514 (2008).

    \bibitem{Beane:2007qr} S.~R.~Beane, W.~Detmold and M.~J.~Savage,
  Phys.\ Rev.\ D {\bf 76}, 074507 (2007).

    \bibitem{Tan:2007bg} S.~Tan,
  arXiv:0709.2530 [cond-mat.stat-mech].

\bibitem{Detmold:2008gh}
  W.~Detmold and M.~J.~Savage,
  Phys.\ Rev.\  D {\bf 77}, 057502 (2008)
  [arXiv:0801.0763 [hep-lat]].

\bibitem{Luu:2008fg}
  T.~Luu,
  PoS {\bf LATTICE2008}, 246 (2008)
  [arXiv:0810.2331 [hep-lat]].

\bibitem{Smigielski:2008pa}
  B.~Smigielski and J.~Wasem,
  Phys.\ Rev.\  D {\bf 79}, 054506 (2009)
  [arXiv:0811.4392 [hep-lat]].

\bibitem{Detmold:2010au}
  W.~Detmold and M.~J.~Savage,
  arXiv:1001.2768 [hep-lat].

\bibitem{Luscher:1986pf}
  M.~L\"uscher,
  Commun.\ Math.\ Phys.\  {\bf 105}, 153 (1986).

\bibitem{mixed}
 W.~Detmold and B.~Smigielski, {\it in preparation}.


\bibitem{Kaufman:1975}
L.~Kaufman,
 BIT \textbf{15}, 49, 1975.

\bibitem{Fleming:2004hs}
G.~T.~Fleming, hep-lat/0403023.

\bibitem{Lepage:2001ym}
  G.~P.~Lepage, {\it et al.},
  Nucl.\ Phys.\ Proc.\ Suppl.\  {\bf 106}, 12-20 (2002).
  [hep-lat/0110175].

\end{thebibliography}
\end{document}